\documentclass[twocolumn,apjl]{emulateapj}

\usepackage{times}

\journalinfo{Accepted for publication in ApJ}
\slugcomment{Accepted for publication in ApJ}

\shorttitle{Spiral arm morphology and black hole mass}
\shortauthors{Seigar et al.}

\begin{document}


\title{Discovery of a relationship between spiral arm morphology and supermassive black hole mass in disk galaxies}


\author{Marc S.\ Seigar\altaffilmark{1, 3}, Daniel Kennefick\altaffilmark{2, 3}, Julia Kennefick\altaffilmark{2, 3} and Claud H.~S.\ Lacy\altaffilmark{2, 3}}
\altaffiltext{1}{Department of Physics \& Astronomy, University of Arkansas at Little Rock, 2801 S.\ University Avenue, Little Rock, AR 72204}
\altaffiltext{2}{Department of Physics, University of Arkansas, 835 West Dickson Street, Fayetteville, AR 72701}
\altaffiltext{3}{Arkansas Center for Space and Planetary Sciences, 202 Old Museum Building, University or Arkansas, Fayetteville, AR 72701}



\begin{abstract}

We present a relationship between spiral arm pitch angle 
(a measure of the tightness of spiral structure) and the 
mass of supermassive black holes (BHs) in the nuclei of disk galaxies. We
argue that this relationship is expected through a combination of 
other relationships, whose existence has 
already been demonstrated. The recent discovery of AGN in bulgeless
disk galaxies suggests that halo concentration or virial mass may
be one of 
the determining factors in BH mass. Taken together with the result
that mass concentration seems to determine spiral arm pitch angle, 
one would expect a relation
to exist between spiral arm pitch angle and supermassive BH mass in disk
galaxies, and we find that this is indeed the case.
We conclude that this relationship
may be important for estimating evolution in BH masses in disk galaxies out 
to intermediate redshifts, since regular spiral arm structure can be seen in 
galaxies out to $z\simeq 1$.

\end{abstract}





\section{Introduction}

Massive black holes (BHs) 
at the centers of galaxies are recognized as a normal, 
perhaps ubiquitous, component of elliptical galaxies and spiral galaxy bulges 
(Kormendy \& Richstone 1995; Barth 2004; Kormendy 2004). 
Indeed, it has been argued that all hot galaxy components (elliptical galaxies
and spiral galaxy bulges) contain central BHs (e.g.,
Magorrian et al.\ 1998). Also, a good correlation has been shown to exist
between the mass of supermassive BHs, $M_{BH}$, and host galaxy mass
or luminosity (Kormendy 1993; Kormendy \& Richstone 1995; Magorrian et al.\ 
1998; Marconi \& Hunt 2003; H\"aring \& Rix 2004). 
A plausible physical framework to 
discuss the connections between BH mass, galaxy formation, 
and quasar evolution was outlined by Richstone et al.\ (1998). Over the
lifetime of HST, the set of galaxies with reliable BH masses has 
grown rapidly, culminating in the 
discovery of a relation between the mass of BHs and the 
stellar velocity dispersion of the bulge/spheroid, $\sigma_c$ 
(Gebhardt et al.\ 2000a; Ferrarese \& 
Merritt 2000). Since the discovery of this $M_{BH}$-$\sigma_c$ relation, 
measurement of lower mass BHs have been made, 
extending mass estimates as low as $10^5 M_{\odot}$ (Barth et al.\ 2004, 
2005; Fillipenko \& Ho 2003; Greene \& Ho 2004) and as high as 
$\sim 3 \times 10^{9} M_{\odot}$ (e.g., Tremaine et al.\ 2002).
This has also allowed investigations of other galaxy properties that 
correlate well with $M_{BH}$, such as bulge light concentration or
S\'ersic index (e.g., Graham \& Driver 2007) and bulge gravitational
binding energy (Aller \& Richstone 2007).

The recent discovery of supermassive BHs in the centers of late-type
galaxies with little or no bulge (Satyapal et al.\ 2007, 2008) suggests that
one of the mechanisms for determining the size (or mass) of centrals BHs 
may be linked to 
the concentration or virial mass of the dark matter halo. This is in 
agreement with Ferrarese (2002), who finds that the mass of the dark matter
halo is linked to BH mass, by comparing maximum rotation velocities, 
$V_{\rm max}$, of disk galaxies with their corrected central velocity
dispersion and finding a good correlation. 
Generally, it can be stated that galaxies with larger bulges tend to be
found in more massive dark matter halos, but this relationship, while
being a good correlation {\em on average}, has a lot of scatter (see, e.g., Ho
2007).
Since spiral arm pitch angle also depends on mass concentration, once 
again this may point to a relation between spiral arm pitch angle and BH mass. 
Such a relation may be important, as it could possibly be 
an indirect means of determining the masses of supermassive BHs in distant 
disk galaxies, and hence the growth of BHs in spirals as a function of look
back time.

The range of supermassive BHs in spirals is less 
than the entire observed range of masses for all galaxy types. For spirals, BH
masses typically range from $\sim10^6 M_{\odot}$ 
(similar to the Milky Way galaxy; Ghez et al.\ 2005; Genzel et al.\ 
2000) to $\sim10^8 M_{\odot}$ (for more massive spiral galaxies like 
M31; e.g., Bender et al.\ 2005). In this letter we show that a correlation
exists between supermassive BH mass and spiral arm pitch angle 
(a measure of the tightness or looseness of spiral arms in disk galaxies).


\section{Observations and Analysis}

Our sample consists of a total of 27 spiral galaxies with BH masses that
have been determined using several different methods. The first 12
galaxies have estimates of their supermassive BHs using direct determinations.
The next 11 galaxies were selected from the sample of Ferrarese (2002) and
the BH masses have been determined using the central velocity dispersion 
of the bulge, $\sigma_c$, and converting to BH mass 
using the relation from Ferrarese (2002). The last 4 galaxies have
lower limits for the BH masses based upon the Eddington limit and have
been taken from the sample of Satyapal et al.\ (2007, 2008). It should be
noted that these lower limits are order of magnitude estimates at best.
Our galaxies consist of those galaxies with known BH masses, with 
Hubble types ranging from Sa to Sm, for which it is possible to measure
spiral arm pitch angle.

\begin{deluxetable*}{llcccccc}
\label{table}
\tablecolumns{8}
\tablewidth{0pc}
\tablecaption{Supermassive BH masses and spiral arm pitch angles for 27 disk galaxies}
\tablehead{
\colhead{Galaxy} & \colhead{Hubble} & \colhead{$\sigma_c$} & \colhead{$M_{BH}$}  & \colhead{Source} & \colhead{$P$}       & \colhead{Wave-}     & \colhead{Archive/} \\
\colhead{Name}            & \colhead{Type}            & \colhead{(km s$^{-1}$)}& \colhead{($M_{\odot}$)}& \colhead{}    & \colhead{(degrees)} & \colhead{band}  & \colhead{Telescope}\\
}
\startdata
\multicolumn{8}{c}{BH estimates from direct measurements}\\
\hline
Milky Way  & SABbc    & $100\pm20$ & $(3.7\pm0.2)\times10^6$ & (1)  & $22\fdg5\pm2\fdg5$ & --- & --- \\ 
Mrk 590    & SAa      & 169        & $(1.6\pm0.9)\times10^7$ & (2)  & $14\fdg3\pm3\fdg0$ & R   & DSS \\
M31        & SAb      & $160\pm10$ & $(1.7\pm0.6)\times10^8$ & (3)  & $7\fdg1\pm0\fdg4$  & --- & --- \\
M33        & SAcd     & $\sim24$   & $<1.5\times10^3$        & (4)  & $42\fdg2\pm3\fdg0$ & R   & DSS \\
IC 342     & SABcd    & $33\pm3$   & $<5.0\times10^5$        & (5)  & $37\fdg1\pm1\fdg3$ & R   & DSS \\
NGC 1068   & SAb      & 151        & $(2.0\pm1.0)\times10^7$ & (6)  & $17\fdg3\pm2\fdg2$ & R   & LCO 2.5-m \\
NGC 3227   & SABb     & 128        & $(3.8\pm3.2)\times10^7$ & (2)  & $13\fdg3\pm0\fdg9$ & R   & DSS \\
NGC 4051   & SABbc    & 88         & $(1.4\pm0.8)\times10^6$ & (2)  & $28\fdg8\pm3\fdg6$ & R   & DSS \\
NGC 4151   & SABab    & 119        & $(1.6\pm0.9)\times10^7$ & (2)  & $20\fdg1\pm0\fdg8$ & R   & Mt.\ Bigelow \\
NGC 4258   & SABbc    & $138\pm18$ & $(3.9\pm0.3)\times10^7$ & (6)  & $8\fdg6\pm2\fdg4$  & R   & Palomar 1.5-m \\
NGC 4395   & SAm      & $<30$      & $10^4-10^5$             & (7)  & $39\fdg8\pm1\fdg6$ & R   & NOT \\
NGC 4593   & SBb      & 124        & $(8.1\pm4.6)\times10^6$ & (2)  & $17\fdg0\pm2\fdg4$ & B   & Lowell 1.1-m \\
\hline
\multicolumn{8}{c}{BH estimates from $\sigma_c$}\\
\hline
NGC 753    & SABbc    & $129\pm18$ & $(2.2\pm0.4)\times10^7$ & (8)  & $13\fdg8\pm1\fdg6$   & I   & INT       \\
NGC 1357   & SAab     & $121\pm14$ & $(1.7\pm0.4)\times10^7$ & (8)  & $18\fdg6\pm2\fdg7$ & R   & LCO 2.5-m \\
NGC 1417   & SABb     & $148\pm18$ & $(4.2\pm0.5)\times10^7$ & (8)  & $10\fdg6\pm1\fdg4$ & R   & LCO 2.5-m \\
NGC 2742   & SAc      & $63\pm28$  & $(8.4\pm4.3)\times10^5$ & (8)  & $34\fdg6\pm3\fdg8$ & J   & Calar Alto 2.2-m \\
NGC 2903   & SBd      & $106\pm13$ & $(9.1\pm1.7)\times10^6$ & (8)  & $17\fdg8\pm0\fdg9$ & R   & DSS       \\
NGC 2998   & SABc     & $113\pm30$ & $(1.2\pm0.4)\times10^7$ & (8)  & $14\fdg4\pm2\fdg5$ & R   & DSS       \\
NGC 3145   & SBbc     & $166\pm12$ & $(7.1\pm0.5)\times10^7$ & (8)  & $9\fdg3\pm1\fdg1$  & R   & LCO 2.5-m \\
NGC 3198   & SBc      & $69\pm13$  & $(1.3\pm0.4)\times10^6$ & (8)  & $30\fdg1\pm2\fdg6$ & J   & Calar Alto 2.2-m \\
NGC 3223   & SAbc     & $163\pm17$ & $(6.5\pm1.0)\times10^7$ & (8)  & $10\fdg7\pm2\fdg0$ & R   & LCO 2.5-m \\
NGC 4321   & SABbc    & $83\pm12$  & $(3.0\pm0.4)\times10^6$ & (8)  & $24\fdg4\pm2\fdg0$ & R   & DSS       \\
NGC 5033   & SAc      & $122\pm9$  & $(1.7\pm0.4)\times10^7$ & (8)  & $14\fdg4\pm0\fdg7$ & R   & DSS       \\
\hline
\multicolumn{8}{c}{BH lower limits from the Eddington Limit}\\
\hline
NGC 3367   & SBc      & ---        & $>1.5\times10^3$        & (9)  & $33\fdg7\pm1\fdg7$ & I   & UNAM 2.1-m\\
NGC 3621   & SAd      & ---        & $>4.0\times10^3$        & (10) & $41\fdg2\pm4\fdg2$ & I   & LCO 2.5-m \\
NGC 3938   & SAc      & 40         & $>1.8\times10^3$        & (9)  & $43\fdg4\pm1\fdg4$ & R   & DSS       \\
NGC 4536   & SABbc    & 84         & $>4.8\times10^3$        & (9)  & $41\fdg4\pm3\fdg9$ & I   & DSS       \\
\enddata
\tablecomments{BH mass estimates and/or velocity dispersions from (1) Ghez et al.\ (2005), (2) Gebhardt et al.\ (2000b), (3) Bender et al.\ (2005), (4) Gebhardt et al.\ (2001), (5) B\"oker et al.\ (1999), (6) Merritt \& Ferrarese (2001), (7) Fillipenko \& Ho (2003), (8) Ferrarese (2002), (9) Satyapal et al.\ (2008), (10) Satyapal et al.\ (2007). The spiral arm pitch angle given for M31 is the average of values taken from Arp (1964) and Braun (1991). The Milky Way pitch angle is from Levine et al.\ (2006). BH estimates from $\sigma_c$ take into account the scatter in the $M_{BH}$-$\sigma_c$ relation.}
\end{deluxetable*}

The source of the
BH mass or velocity dispersion of the galaxies in our sample is given in 
Table 1. 
Images of 25 of these galaxies were then downloaded
from various archives, also listed in Table 1. The images were then used to 
determine spiral arm pitch angles (see Table 1). 
The remaining 2 galaxies are the Milky Way
and M31, for which spiral arm pitch angles were 
taken from
the literature. Levine et al.\ (2006) measure the spiral arm pitch angle of
the Milky as $P=22\fdg5\pm2\fdg5$ from neutral hydrogen observations. For
M31, Arp (1964) measured a pitch angle $P=7\fdg4$ and Braun (1991) measured a 
pitch angle $P=6\fdg7$. Here we adopt the average of these two measurements 
for the spiral arm pitch angle of M31.
The remaining spiral arm pitch angles were measured using a two-dimensional 
fast Fourier transformation (Schr\"oder et al.\ 1994), assuming 
logarithmic spirals.

The range of radii over which the Fourier fits were applied were selected to
exclude the bulge or bar (where no information about the arms exists) and 
to extend out to the outer limits of the arms in our images. 
Pitch angles were
then determined from peaks in the Fourier spectra, as this is the most powerful
method to find periodicity in a distribution (Consid\`ere \& Athanassoula 1988;
Garcia-Gomez \& Athanassoula 1993). The radial range over which the Fourier
analysis was performed was chosen by eye and is probably the dominant source
of error in the calculation of pitch angles. As a result, three radial ranges
were chosen for each galaxy, and a mean pitch angle and standard error 
calculated for every object.

The images were first deprojected to face-on. Mean uncertainties of position
angle and inclination as a function of inclination were discussed by 
Consid\`ere \& Athanassoula (1988). For a galaxy with low inclination, there
are clearly greater uncertainties in assigning both a position angle and an
accurate inclination. These uncertainties are discussed by Block et al.\ (1999)
and Seigar et al.\ (2005), who take a galaxy with low inclination 
($<30^{\circ}$) and one with high inclination ($>60^{\circ}$) and vary the
inclination angle used in the correction to face-on. They find that for the
galaxy with low inclination, the measured pitch angle remains the same. 
However, the measured pitch angle for the galaxy with high inclination varies 
by 10\%. For galaxies with inclination $i>60^{\circ}$ we take into 
account this uncertainty. Our deprojection method assumes that spiral galaxy
disks are intrinsically circular.

\section{Results and Discussion}

Figure 1 shows a plot of supermassive BH mass versus spiral arm pitch angle.
As expected, we find a good correlation between BH mass and pitch angle.
Pearson's linear correlation coefficient is 0.95 and the significance level at 
which the null hypothesis of zero correlation is disproved is 3$\sigma$. 
This means that the masses of BHs in the nuclei of disk galaxies can be
determined directly from a measurement of their spiral arm morphology, in
particular the spiral arm pitch angle.

\begin{figure}
\label{mb}
\plotone{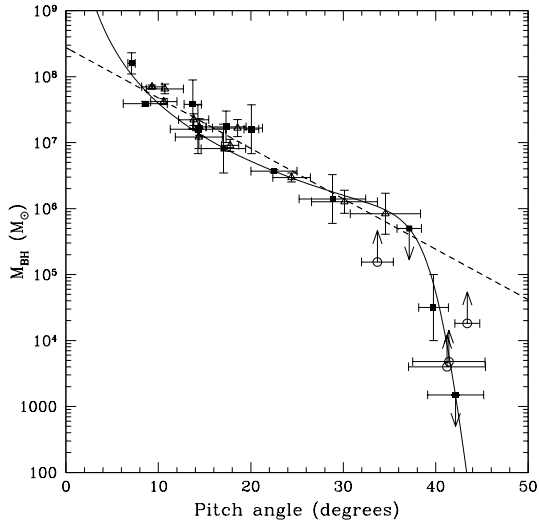}
\caption{Plot of supermassive BH mass as a function of spiral arm pitch angle. Solid squares represent galaxies with direct measurements of their central BH mass, triangles represent galaxies for which the BH mass has been determined from the bulge central velocity dispersion and open circles represent lower limits on BH masses from the Eddington limit. The solid line represent the best fit analytical model describing the relationship between BH mass and spiral arm pitch angle, as described by equation 2. The dashed line is a linear best fit for galaxies with pitch angles $P<38^{\circ}$.}
\end{figure}

The relationship shown in Figure 1 can be explained as follows. 
It has been shown that
spiral arm pitch angle is related to the rotation curve shear of disk galaxies
(Seigar, Block \& Puerari 2004; Seigar et al.\ 2005, 2006). The rotation curve 
shear is a measure of the
slope of the rotation curve and is therefore directly related to the overall
mass concentration (see Seigar et al.\ 2006 for a definition of rotation curve
shear). Furthermore, the recent discovery of AGN in late-type
bulgeless galaxies (Satyapal et al.\ 2007, 2008) suggests that BH mass may
be more closely related to mass concentration than bulge mass. Satyapal
et al.\ (2008) suggest that the dark matter halo virial mass (which is
related to the dark matter halo concentration; see e.g., van den Bosch et al.\
2007) is the key quantity in determining BH mass. Because the BH mass and
spiral arm pitch angle are both related to mass concentration, it is no
surprise that we find that they are related to each other. However, it should
be noted that to truly determine if mass concentration is the main determinant
of BH mass is the subject of a more detailed study.

\begin{figure}
\label{sig}
\plotone{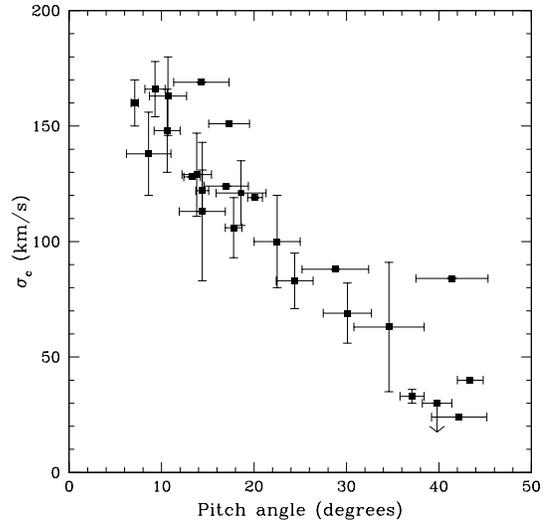}
\caption{Plot of bulge central velocity dispersion as a function of spiral arm pitch angle, showing a good correlation.}
\end{figure}

For galaxies with the most open spiral structure (with
pitch angles $>40^{\circ}$) the slope in the relationship appears to change 
sharply.
The most open spiral structure in disk galaxies has typical pitch angles
between 40$^{\circ}$ and 45$^{\circ}$ (e.g., Seigar \& James 1998; Block et
al.\ 1999). These galaxies have small bulges, if any bulge at all
(Kennicutt 1981; de Jong 1996; Seigar \& James 1998). As a result
the BH masses in galaxies with very open spiral arms may very quickly
approach zero. In this study, the galaxy with the most open spiral structure
is M33, with a pitch angle, $P=42\fdg2\pm3\fdg0$. This galaxy has a Hubble
classification of SAcd, and is often referred to as being a bulgeless galaxy
(e.g., Stephens \& Frogel 2002).
Furthermore, it has the lowest measured mass of any nuclear BH ever measured,
with an upper limit of 1500 $M_{\odot}$ (Gebhardt et al.\ 2001). If these
bulgeless galaxies have nuclear BH masses that are very low, or
possibly approaching zero, the sharp change in the slope of the relation
above pitch angles of 40$^{\circ}$ is probably due to a cut-off in the
expected range of black hole masses for disk galaxies. It should also be
noted, that this change in slope is currently driven by just 5 galaxies with
pitch angles, $P>38^{\circ}$. If
a larger sample were available it may be demonstrated that the correlation
simply breaks down for galaxies with the most open spiral structure. For this
reason we have used a linear relation to describe galaxies with pitch angles
$P<38^{\circ}$ (Figure 1; dashed line). The equation of this line is,
\begin{equation}
\log_{10}{M_{BH}}=(8.44\pm0.10)-(0.076\pm0.005)P.
\end{equation}

The smallest pitch angles typically seen are
in the range $7^{\circ}-10^{\circ}$ (Block et al.\ 1999; Seigar 2005). Such
galaxies have the largest bulges and subsequently the most massive nuclear
BHs (for spirals). 
The galaxy with the largest black hole in this study is M31, with a mass
of $\sim1.7\times10^8 M_{\odot}$ (Bender et al.\ 2005). M31 is classified as 
SAb, but it is likely mis-classified due to its high inclination, and most
likely has a larger bulge and/or bar (e.g., Beaton et al.\ 2007).

Assuming that the break in the relation between SMBH mass and spiral arm
pitch angle in Figure 1 is real, we have also used a double power-law model 
(similar to the Nuker law of Lauer et al.\ 1995) to describe the relation
(Figure 1; solid line). Such a model is written:
\begin{equation}
M_{BH}=2^{\left(\frac{\beta-\gamma}{\alpha}\right)}M_{BH_b}\left(\frac{P_b}{P}\right)^{\gamma}\left[1+\left(\frac{P}{P_b}\right)^{\alpha}\right]^{\left(\frac{\gamma-\beta}{\alpha}\right)}
\end{equation}
where $\gamma$ measures the slope of the power law for low pitch angles, 
$\beta$ is the slope of the power law for large pitch angles, $P_b$ is the 
transition from low pitch angles to high pitch angles, $\alpha$ governs the 
sharpness of the transition and $M_{BH_b}$ is the black hole mass for a pitch 
angle $P_b$. The solid line in Figure 1 is the best fitting model described by
equation 2. The scatter with respect to this best-fit is 0.3 dex. 
For the best fit we find values for the parameters, $\alpha=23.5$,
$\beta=126.1$, $\gamma=2.92$, $M_{BH_b}=1.72\times10^4 M_{\odot}$ and
$P_b=40\fdg8$. Also, it can be seen that as the pitch angle
approaches zero, that the black hole mass increases rapidly. However, it 
should be noted that the relationship only applies to spiral galaxies, and
is therefore not applicable below pitch angles of $\sim7^{\circ}$, i.e. similar
to the pitch angle of M31, since this is the tighest pitch angle typically
seen in disk galaxies.

In Figure 2 we show a plot of bulge central velocity dispersion,
$\sigma_c$, as a function of spiral arm pitch angle. Not surprisingly, a 
good correlation is seen. This is simply a result of the fact that $\sigma_c$
is also an indicator of the mass of central BHs (Gebhardt et al.\ 2000a;
Ferrarese \& Merritt 2000). In this case,
Pearson's linear correlation coefficient is 0.92 and the significance level at 
which the null hypothesis of zero correlation is disproved is also 3$\sigma$.

We suggest that two of the factors in the determination of supermassive
BH mass could be either dark matter halo concentration
($c=R_{\rm vir}/R_{\rm s}$, where $R_{\rm vir}$ is the virial radius and 
$R_{\rm s}$ is the dark matter density profile scale radius; Navarro et al.\ 
1997) or the dark matter halo virial mass (see Ferrarese 2002). One of 
the determining factors
of spiral arm pitch angle is central mass concentration, which in turn is
related to dark matter halo concentration (e.g., Seigar et al.\ 2006). 
Furthermore, spiral galaxies with larger halo concentrations are more likely 
to have larger stellar bulges in which more massive BHs can reside. Finally, 
the stellar velocity dispersion of the bulge, $\sigma$, is an indicator of the 
dynamical mass within the radius at which it is measured. This is
also related to halo concentration, which in turn is related to halo
virial mass (e.g., van den Bosch et al.\ 2007). 
It is therefore possible not only to explain the relationship between 
spiral arm pitch angle and supermassive BH mass (presented here), but also 
the relation between bulge mass and
supermassive BH mass (Magorrian et al.\ 1998) and the $M_{\rm BH}$-$\sigma$
relation (Gebhardt et al.\ 2000a; Ferrarese \& Merritt 2000). However, we note
that given the current evidence, it is impossible to tell if dark matter
halo concentration is the main determinant of supermassive BH mass, or if
some other quanitity is also important. To determine if the dark matter halo
does affect BH mass is the subject of a more detailed study.

Finally, the importance of the relationship between spiral arm pitch angle
and supermassive BH mass lies in the ability to detect regular spiral structure
out to intermediate redshifts. For example, Elmegreen et al.\ (2004) clearly
show that disk galaxies at $z\simeq 1$ can exhibit grand design spiral 
structure. If this is true, then it may be possible to estimate the growth
of supermassive BHs in disk galaxies as a function of look back time, just
from imaging data and a measurement of spiral arm pitch angle. In such a
study, the accuracy for predicting the BH mass for an individual galaxy
is $\pm$30\%, based on the scatter in Figure 1 and the typical errors
associated with estimating spiral arm pitch angles. Taking into account
that spiral arm pitch angles for more distant galaxies will be measured with 
less certainty, it is likely that we will be able to determine BH masses to
within $\pm$50\%. It is therefore important to note that the 
spiral arm pitch
angle versus BH mass relation should be applied to a large sample of nearly
face-on galaxies, in which case a statistical analysis of the evolution of BH
masses in disk galaxies can be determined. Given the quantity of deep
HST imaging of blank fields that is now available, it seems that the time is 
ripe to perform such as study. For example, we estimate that $\sim$700
galaxies are available for a study of spiral structure 
from the Great Observatories Origins Deep
Survey fields alone (see Giavalisco et 
al.\ 2004 for a description of this survey). It therefore should now be 
possible to determine, statistically, the evolution of SMBHs in disk galaxies,
based upon spiral arm morphology.

\acknowledgments

Support for this work was provided by the Arkansas NASA EPSCoR program.
This research has made use of the NASA/IPAC Extragalactic Database (NED) 
which is operated by the Jet Propulsion Laboratory, California Institute 
of Technology, under contract with NASA.
The authors wish to thank the anonymous referees whose comments improved
the content of this letter.

\end{document}